\documentclass{article}

\usepackage{arxiv}
\pdfoutput=1 
\usepackage{ifpdf}
\usepackage[utf8]{inputenc} % allow utf-8 input
\usepackage[T1]{fontenc}    % use 8-bit T1 fonts
\usepackage{hyperref}       % hyperlinks
\usepackage{url}            % simple URL typesetting
\usepackage{booktabs}       % professional-quality tables
\usepackage{amsfonts}       % blackboard math symbols
\usepackage{nicefrac}       % compact symbols for 1/2, etc.
\usepackage{microtype}      % microtypography
\usepackage{lipsum}		% Can be removed after putting your text content
\usepackage{amsmath}
\usepackage{amssymb}
\usepackage{float}
\usepackage{color}
\usepackage{colortbl}

\usepackage{graphicx}
\usepackage{subcaption}
\usepackage{caption}
\usepackage{hyperref}
\title{VLC-Based Networking: Feasibility and Challenges}

\author{
  Alain R.~Ndjiongue\thanks{Alain R. Ndjiongue is with the Faculty of Engineering and the Built Environment, University of Johannesburg.} \\
  \texttt{ndjiongue@gmail.com} \\
   \And
 Telex M. N. Ngatched \thanks{Telex M. N. Ngatched is with the Faculty of Engineering and Applied Science, Memorial University.} \\
  \texttt{tngatched@grenfell.mun.ca} \\
 \And
 Octavia A. Dobre \thanks{Octavia A. Dobre is with the Faculty of Engineering and Applied Science, Memorial University.} \\
 \texttt{odobre@mun.ca} \\
  \And
  Ana G. Armada \thanks{Ana G. Armada is with the Signal Theory and Communications Department, Universidad Carlos III de Madrid.} \\
  \texttt{anagar@ing.uc3m.es} \\
}

\date{}

\begin{document}
\maketitle

\begin{abstract}
Visible-light communication (VLC) has emerged as a prominent technology to address the radio spectrum shortage. It is characterized by the unlicensed and unexploited high bandwidth, and provides the system with cost-effective advantages because of the dual-use of light bulbs for illumination and communication and the low complexity design. It is considered to be utilized in various telecommunication systems, including 5G, and represents the key technology for light-fidelity. To this end, VLC has to be integrated into the existing telecommunication networks. Therefore, its analysis as a network technology is momentous. In this article, we consider the feasibility of using VLC as a network technology and discuss the challenges related to the implementation of a VLC-based network, as well as the integration of VLC into existing conventional networks and its inclusion in standards.
\end{abstract}

% keywords can be removed

\section{Introduction}
Visible light communication (VLC) is a data transmission technology which exploits the light beam as a communication medium. It is a variant of optical wireless communications, which uses light-emitting diodes (LEDs) as antennas and is characterized by short transmission range. In the indoor environment, VLC provides both data transmission and illumination \cite{7239528}. Its main applications include light-fidelity (Li-Fi), indoor positioning, as well as vehicle-to-vehicle and infrastructure-to-vehicle communications. It has recently been demonstrated that using laser diodes (LDs), VLC can also be utilized in access networks \cite{2222018}. The efficient utilization of VLC for such applications requires its analysis as a network technology, which is the focus of this article. \par The interest in the VLC technology can be evidenced by the growing number of research works in the literature \cite{7239528, 6675792, 2016222, 7402263, 6935084} (and references therein), the prototypes and field trials proposed by researchers and industry, and by the standardization activities. Following these efforts, the effective deployment is seen as the next key step. Several challenges are to be overcome when practically deploying communication systems using VLC in both simplex and duplex configurations. Among these challenges, we underline the fact that the return path in a VLC-based communication system may use a different technology, such as radio frequency (RF), power line communication, fiber optics, and free space optical. If using VLC, the return path can be based on a different wavelength. Another important challenge is the illumination coverage. The VLC transmission range is limited, as it is imposed by the lighting coverage of the source, which is short by nature \cite{8569020}. LEDs and LDs are the two main types of light sources considered in VLC. Compared to LEDs, whose lighting range is only a couple of hundred meters, LDs provide an illumination coverage in the order of a couple of kilometers. On the other hand, since only positive and real signals can be successfully transmitted in VLC systems, the modulation schemes that can be employed should produce a non-complex and asymmetric signal. \par The fact that light cannot go through opaque objects confers a security aspect to VLC. However, it compels the indoor VLC-based network to be confined in a room. Consequently, VLC can be used in optical wireless personal area network (OWPAN) and several OWPANs may be linked using VLC to form a wide OWPAN (W-OWPAN). In OWPANs with multiple access points (APs) and mobile user devices (UDs), handover is one of the most challenging problems \cite{8101446}. \par While there are many proposals for VLC transmitter and receiver, the success of the technology depends on the ability to create networks with them and facilitate the deployment. This article discusses the use of VLC as a network technology and its integration into existing networks. It describes VLC-based networking and presents topologies and network structures. It also examines the main types of VLC-based W-OWPAN and introduces connectivity configurations. The article proposes an architecture for VLC-based networks, highlights challenges related to VLC-based network design and implementation, and underlines the recommendations of IEEE on VLC networking. Finally, the future of networking using VLC is discussed.   
\section{VLC-based Networking}
This section introduces VLC-related networks, highlights link configurations and types of networks, and provides an overview of the communication protocols for VLC-based networks. 
\subsection{VLC Connectivity Configurations}
\begin{figure}
	\centering
	\begin{subfigure}{0.70\textwidth}
		\includegraphics[width=\textwidth]{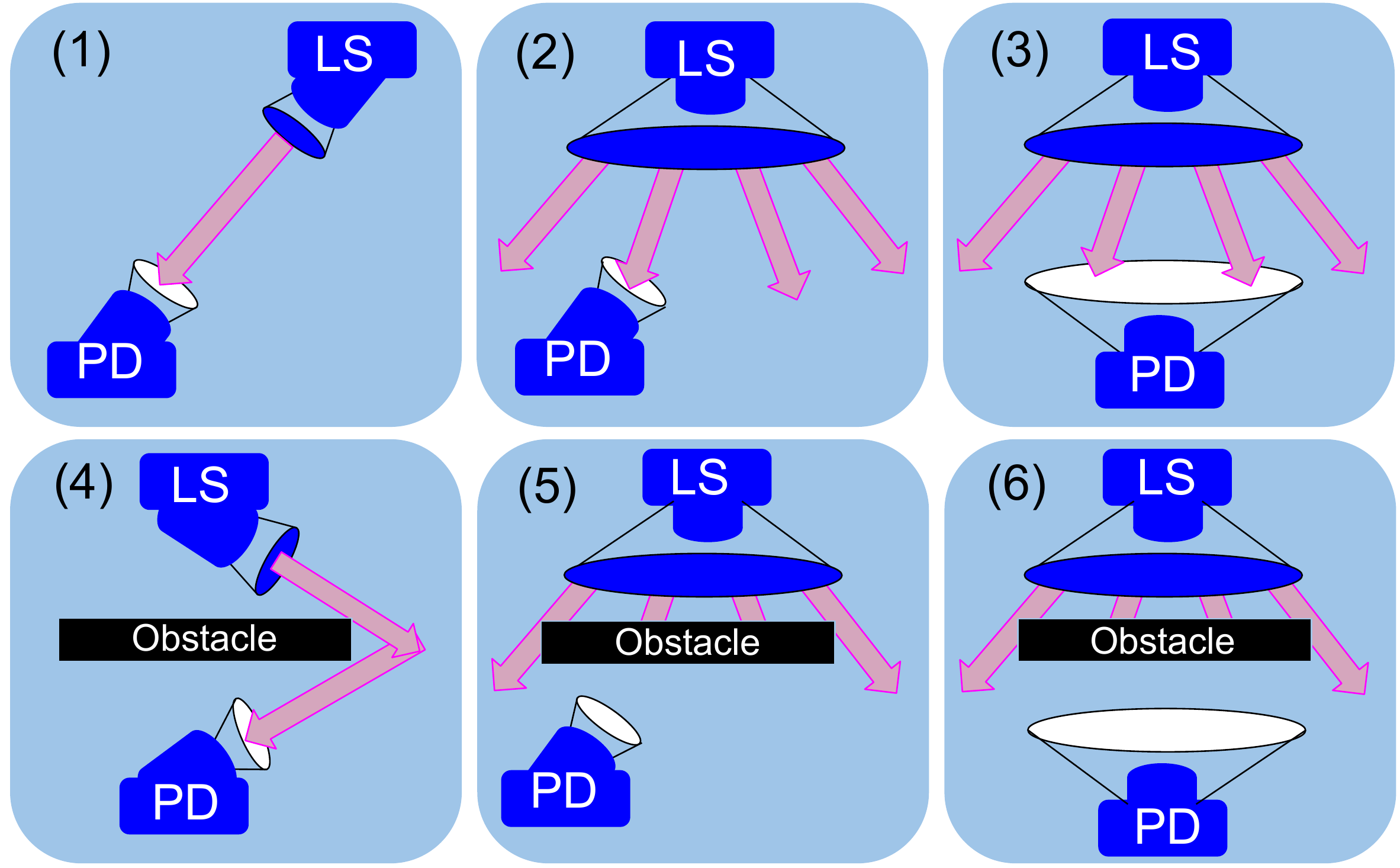}
		\caption{Typical VLC connectivity configurations.}
		\label{Fig:10}
	\end{subfigure}
	\begin{subfigure}{0.35\textwidth}
		\includegraphics[width=\textwidth]{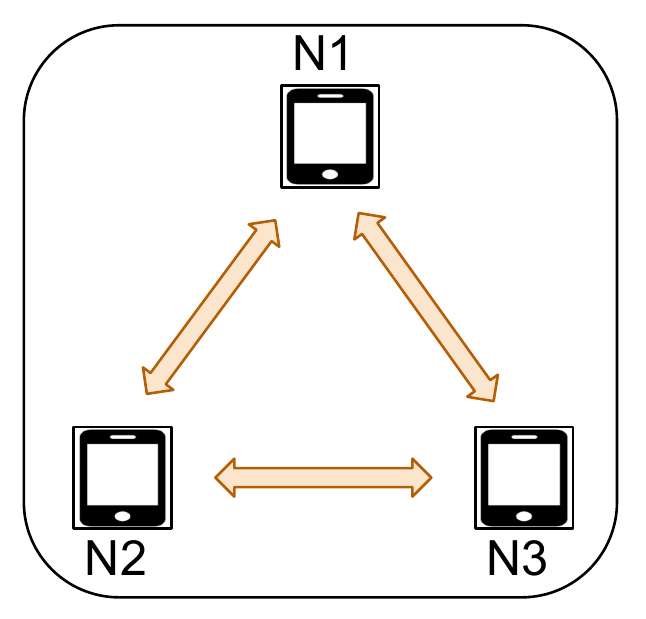}
		\caption{VLC-based P2P network.}
		\label{Fig:11}
	\end{subfigure}
	\begin{subfigure}{0.35\textwidth}
		\includegraphics[width=\textwidth]{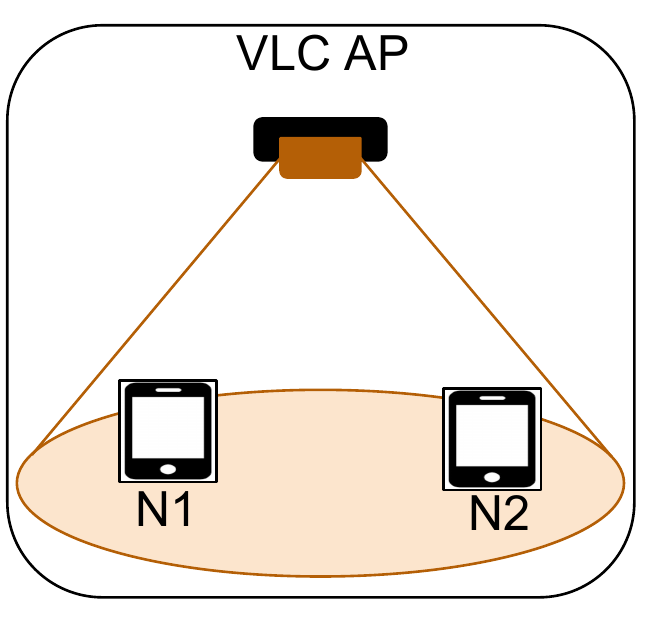}
		\caption{VLC-based OWPAN.}
		\label{Fig:12}
	\end{subfigure}
	\caption{Illustration of VLC-based PANs. LS: light source.}
	\label{Fig:1}
\end{figure}

Two main topologies, defined by the shape of the light beam, govern the transmission in VLC-based networks \cite{2018111}: point-to-point (P2P) and point-to-multipoint (P2MP) topologies. In the P2P topology, the light source, which has a narrow beam with a small divergence angle, is only exploited for communication and cannot be used for illumination. In the P2MP topology, also defined as AP-based VLC network, the beam has a diffuse or a quasi-diffuse shape. In this case, the light source serves as communication antenna and illumination device simultaneously. These topologies, together with the connectivity mode, critically characterize the design of a VLC-based network. In these topologies, six connectivity scenarios can be defined based on the link between the light source and the photodetector (PD), as depicted in Fig.~\ref{Fig:1}. Scenarios (1), (2), and (3) represent line-of-sight cases, whereas Scenarios (4), (5), and (6) utilize non-line-of-sight links; the latter respectively correspond to the former when an obstacle constrains the message signal to reach the UD through reflected rays. \par The light source and the UD may be connected based on the P2P configuration (Fig.~\ref{Fig:11}), where the VLC receiver is made up of a narrow field-of-view [Fig.~\ref{Fig:10}, Scenario (1)]. Using the P2MP configuration (Fig.~\ref{Fig:12}), the receiver may have a narrow field-of-view [Fig.~\ref{Fig:10}, Scenario (2)], or a wide field-of-view [Fig.~\ref{Fig:10}, Scenario (3)]. \par VLC-based networks utilizing APs, i.e., OWPAN, may have two variants: (\textit{i}) the conventional distributed architecture, where each light source contains both the electrical front-end and the optical front-end, and thus, performs both baseband processing and power allocation. Each VLC AP serves as a unique optical transceiver. It receives the message to be re-transmitted from the backbone network and proceeds with the re-transmission; (\textit{ii}) the centralized light access network architecture, in which all VLC APs computational resources are grouped into a unique centrally managed pool \cite{4442017}. Here, the AP contains only an optical front-end, and thus, is only in charge of connecting the light source to both the communication and electrical signals. The centralized light access network architecture decreases the associated cost and complexity of each VLC light source for an efficient implementation, and also reduces interference from neighboring cells. 
\subsection{VLC-based Communication Protocol}
VLC-based networks use a communication protocol which is adapted from the generic Open Systems Interconnection protocol model and utilizes the layered-architecture of the Transmission Control Protocol/Internet Protocol (TCP/IP) model. When compared to protocols developed for RF-based networks, the communication protocol for VLC-based networks mainly differs on the two bottom layers. This difference is crucial in the design of VLC-based networks because the antenna used in the VLC technology (LED/LD) may also be exploited for illumination. Consequently, the design of a VLC-based network card, which includes an LED/LD or a PD, is different from that of the conventional network card. The physical (PHY) layer, on the transmitter side, decides on the bit representation across the networking medium and on the illumination level, while on the receiver side, it is in charge of detecting the message-signal. An example of a protocol for VLC-based networks, referred to as decoupled TCP, is proposed in \cite{1122018}. While this protocol utilizes the layered-architecture of the TCP/IP model, it considers a unidirectional link between VLC APs and UDs in an aggregate VLC-based network.
\subsection{Classification of VLC-based Networks}
VLC-based networks are defined by five main characteristics: (\textit{i}) the transmission distance, which determines if the system is relayed or non-relayed; (\textit{ii}) the need of a return path, which makes the network duplex or simplex; (\textit{iii}) the technology used in the return path, which defines whether the network is standalone or aggregate; (\textit{iv}) the category of UDs and protocols, which specifies whether the network is homogeneous or heterogeneous; (\textit{v}) the use of single or multiple channels to transmit data. These network types are presented in Fig.~\ref{Fig:3} and discussed in the following.
\subsubsection{Relayed vs. Non-relayed VLC-based Networks}
\begin{figure}
	\centering
	\begin{subfigure}{0.63\textwidth}
		\includegraphics[width=\textwidth]{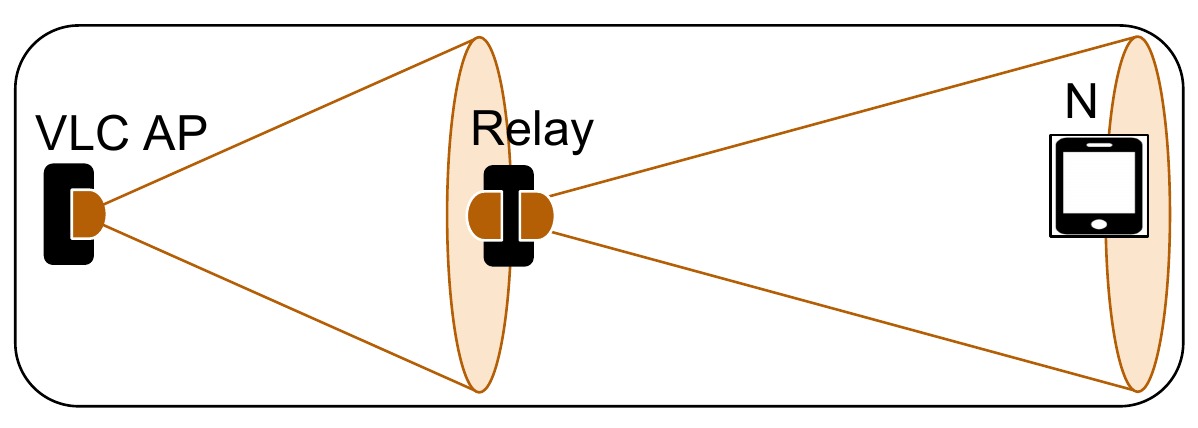}
		\caption{Relayed VLC-based network.}
		\label{Fig:30}
	\end{subfigure}
	\begin{subfigure}{0.63\textwidth}
		\includegraphics[width=\textwidth]{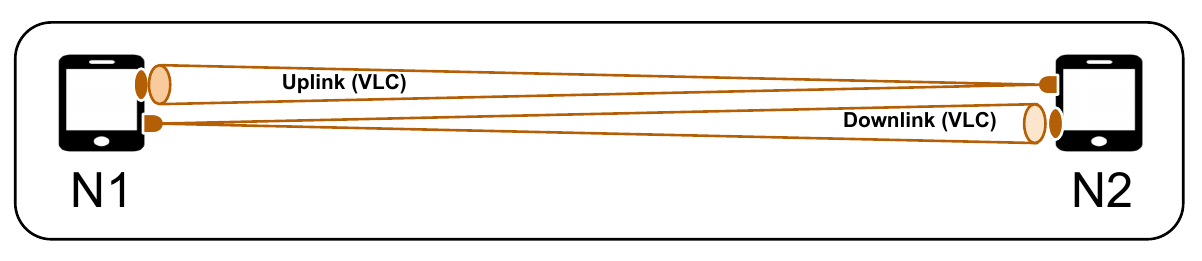}
		\caption{Standalone VLC-based network highlighting both uplink and downlink channels, which are VLC channels.}
		\label{Fig:21}
	\end{subfigure}
	\begin{subfigure}{0.63\textwidth}
		\includegraphics[width=\textwidth]{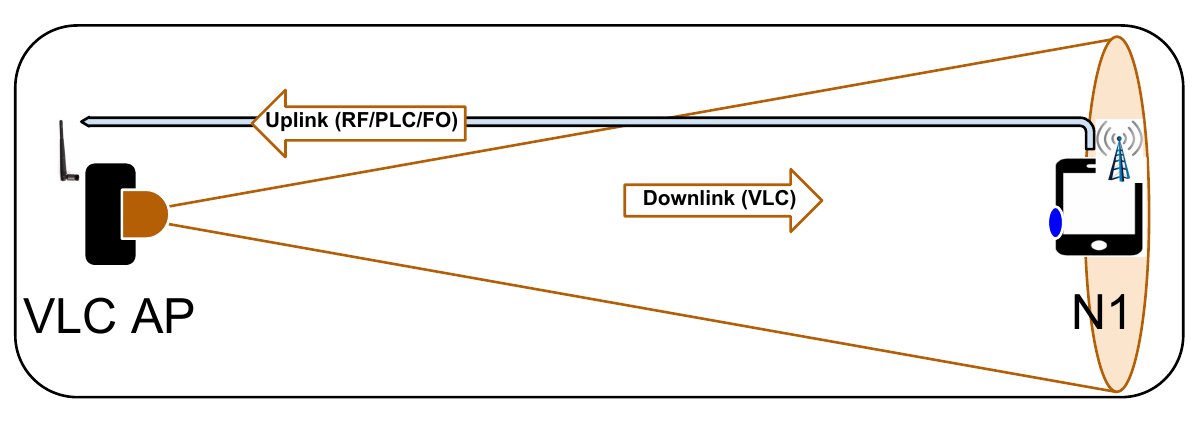}
		\caption{Aggregate VLC-based network: Downlink channel is VLC based and uplink uses a different technology.}
		\label{Fig:22}
	\end{subfigure}
	\begin{subfigure}{0.30\textwidth}
		\includegraphics[width=\textwidth]{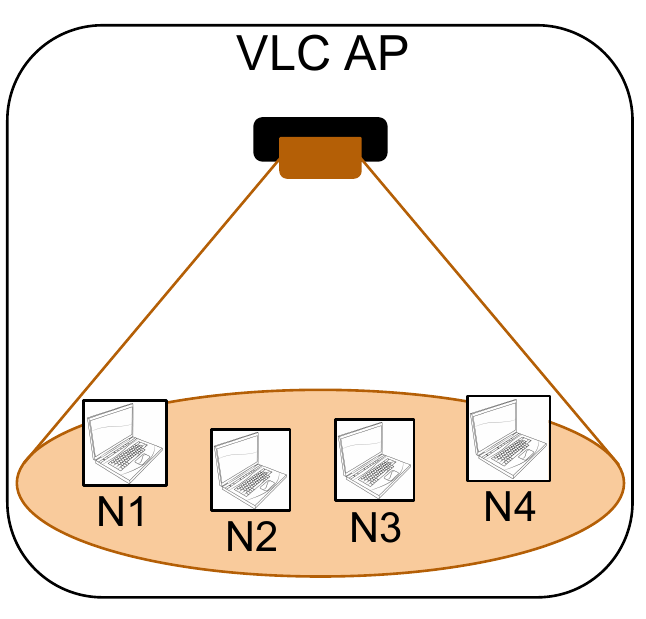}
		\caption{Homogeneous VLC-based network.}
		\label{Fig:31}
	\end{subfigure}
	\begin{subfigure}{0.30\textwidth}
		\includegraphics[width=\textwidth]{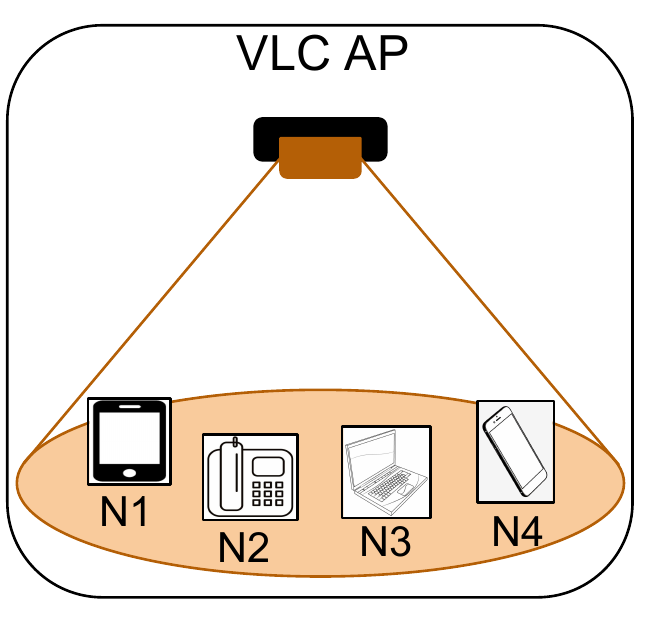}
		\caption{Heterogeneous VLC-based network.}
		\label{Fig:32}
	\end{subfigure}
	\begin{subfigure}{0.63\textwidth}
		\includegraphics[width=\textwidth]{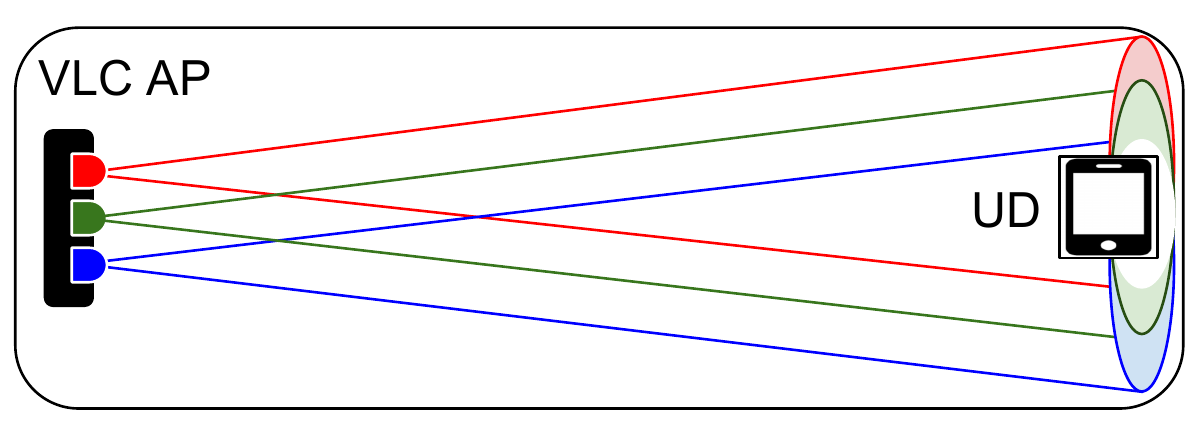}
		\caption{Multi-channel VLC-based network.}
		\label{Fig:33}
	\end{subfigure}
	\caption{Relayed, standalone, aggregate, homogeneous, heterogeneous, and multi-channel VLC-based networks.}
	\label{Fig:3}
\end{figure}
To create a network with VLC, the data to be transmitted may be generated locally or remotely, depending on the transmitter-receiver distance. Hence, networks with VLC can be classified as relayed or non-relayed networks. For the latter, the transmission distance is within the lighting range of the light source. Here, a relay is not needed because the receiver is situated at the correct distance under the transmitter's light beam. \par On the contrary, if the transmission distance is outside the illumination coverage of the light source, or if the light is severely attenuated at the receiver, as illustrated in Fig.~\ref{Fig:30}, then the transmitted message is first sent to a relay, which then forwards it to the recipient \cite{7180320}. In this case, the transmission may be based on IEEE 802.15.7 (VLC), IEEE 802.3 (Ethernet), IEEE 802.11 [wireless fidelity (Wi-Fi)], IEEE 1901 (power line communications), etc. An example of a relayed VLC-based network is the Li-Fi, in which the original message is transmitted to the VLC AP via RF, fiber optics or any other data communication technology. \par Note that cooperative communication techniques which have been widely used in RF systems to improve reliability, can also be employed in relayed VLC-based networks \cite{7180320}. Traditional cooperative techniques, such as amplify-and-forward and decode-and-forward can be adopted. More spectral-efficient relaying techniques, such as incremental decode-and-forward and incremental selective decode-and-forward, can also be used.
\subsubsection{Simplex vs. Duplex VLC-based Networks}
In numerous VLC applications such as infrastructure-to-vehicle communications, the return path is not required. The message is transmitted from one node to another and not vice-versa. This type of links, also known as simplex links, present the advantage of low-complexity and ease of implementation. However, if the application requires a return path, Li-Fi for example, then the link is said to be duplex. Duplex links use both downlink and uplink traffic and have higher design complexity. Technologies for return path (uplink traffic) can be any other data communication technology, such as those listed for relayed VLC-based networks. Simplex and duplex VLC links can be combined with relayed and non-relayed networks.
\subsubsection{Standalone vs. Aggregate VLC-based Networks}
Standalone and aggregate VLC-based networks are based on a duplex link, but differ on whether or not the same technology is used in both downlink and uplink connections. \par In the former, both downlink and uplink use the VLC technology \cite{2222018}. As shown in Fig.~\ref{Fig:21}, the duplex connection between two nodes using VLC will be more efficient if the two (downlink and uplink) light beams do not interfere, and correspond to two P2P links. If the light sources are also used for illumination, then numerous interference signals and significant path loss are expected. This would occur even if the links use different wavelengths. \par For the latter, VLC is used for downlink and a different technology is  employed in uplink. This configuration is illustrated in Fig.~\ref{Fig:22}, where the light source is employed to guide the downlink traffic, and serves as an illumination device as well. This configuration has been investigated in the literature in the deployment of Li-Fi \cite{7402263} and the practical implementation of 5G \cite{7224733}, in both its enhanced mobile broadband and massive machine type communications slices. The former uses Wi-Fi, whereas the latter uses celular to complement the VLC. A variance of aggregate VLC-based networks can be obtained by connecting the UD to the router using a duplex VLC in parallel with another technology, resulting in two independent connections between the router and the UD. As an example, VLC and RF can be combined such that the UD is connected to the router via both VLC AP and RF AP in both downlink and uplink \cite{7224733}. This helps to manage the traffic on the RF channel and release its spectrum during congestion time. 
\subsubsection{Homogeneous vs. Heterogeneous VLC-based Networks}
The concept of homogeneity/heterogeneity is related to the similarity between UDs and their connectivity protocols, and is generally associated with AP-based networks. In a homogeneous VLC-based network, several similar UDs are connected to the VLC AP using analogous connectivity protocols, whereas in a heterogeneous VLC-based network, different types of UDs with different connectivity protocols are linked to the VLC AP. These two types of VLC-based networks are illustrated in Figs.~\ref{Fig:31} and \ref{Fig:32}. Note that the concept of VLC heterogeneous networks is to be contrasted from aggregate VLC-based networks, which are also seen and defined by some researchers as heterogeneous networks due to the use of different technologies.
\subsubsection{Multi-channel VLC-based Networks}
A multi-channel VLC-based network can be implemented using a class of light-sources that exploits multiple wavelengths to generate light \cite{6809190}. In contrast to single channel VLC-based networks, the beam in multi-channel VLC-based networks is formed by several single light beams, as illustrated in Fig.~\ref{Fig:33} for a single user. Typically, such networks use the color-shift-keying scheme \cite{6809190} or may exploit the multiple-input multiple-output technique \cite{8063317} to transmit data. In both cases, considering a single user, the data rate is improved when compared with single channel systems. In the case where multiple users are connected to the AP, a multi-channel VLC-based network also helps to multiplex the users.
\section{Proposed architecture for VLC-based Networks}
\begin{figure}[h!]
	\centering
	\begin{subfigure}{0.96\textwidth}
		\includegraphics[width=\textwidth]{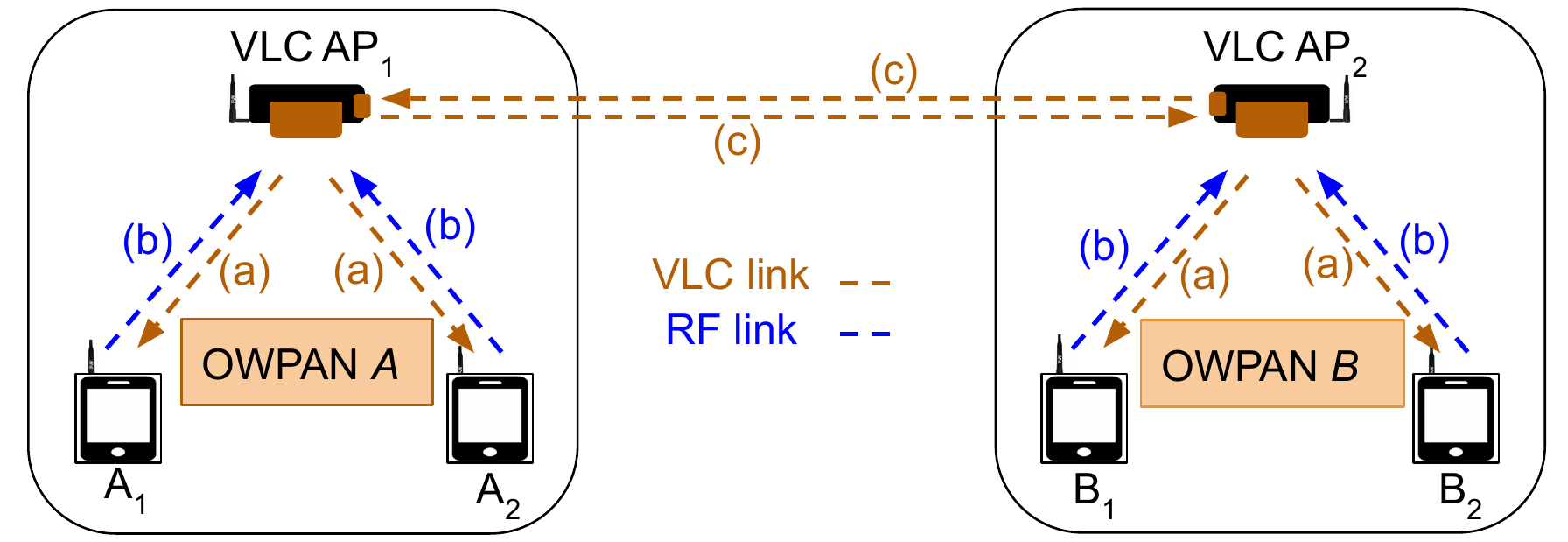}
		\caption{A proposed VLC-based W-OWPAN.}
		\label{Fig:40}
	\end{subfigure}
	\begin{subfigure}{0.48\textwidth}
		\includegraphics[width=\textwidth]{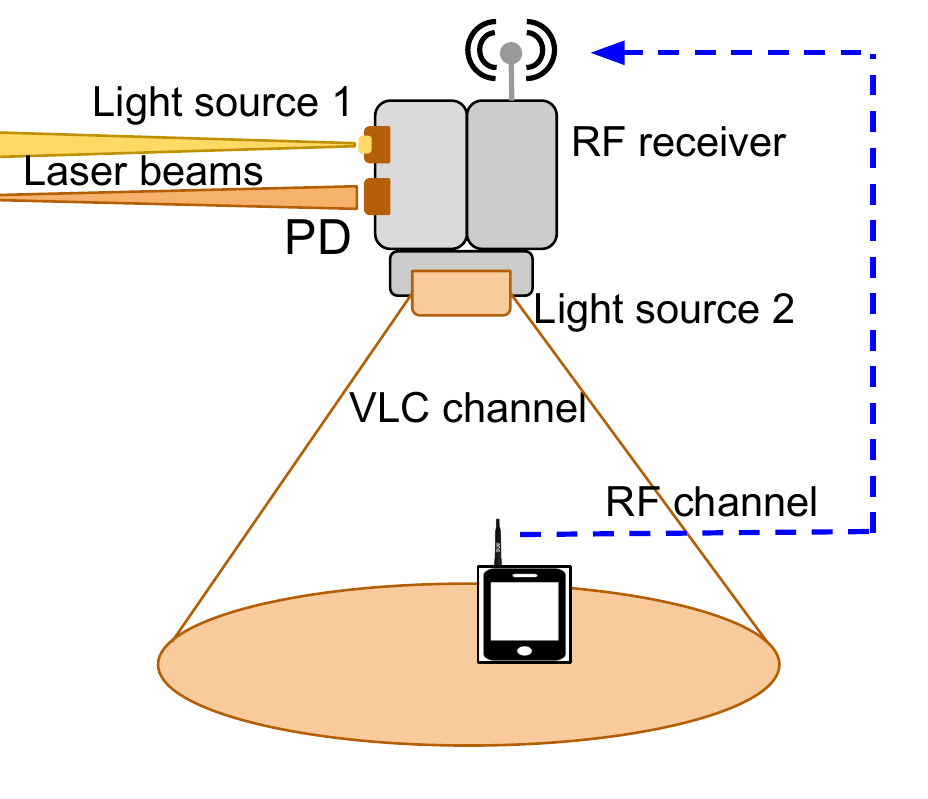}
		\caption{Structure of VLC AP with communication links.}
		\label{Fig:400}
	\end{subfigure}
	\begin{subfigure}{0.48\textwidth}
		\includegraphics[width=\textwidth]{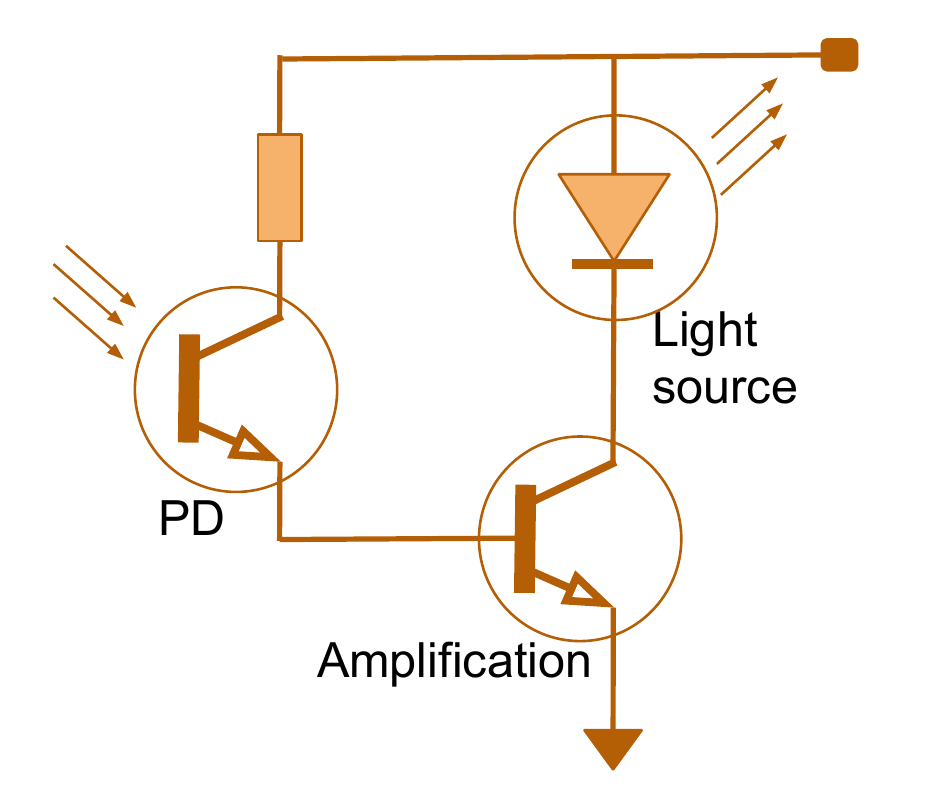}
		\caption{Example of amplification circuit.}
		\label{Fig:401}
	\end{subfigure}
	\begin{subfigure}{0.96\textwidth}
		\includegraphics[width=\textwidth]{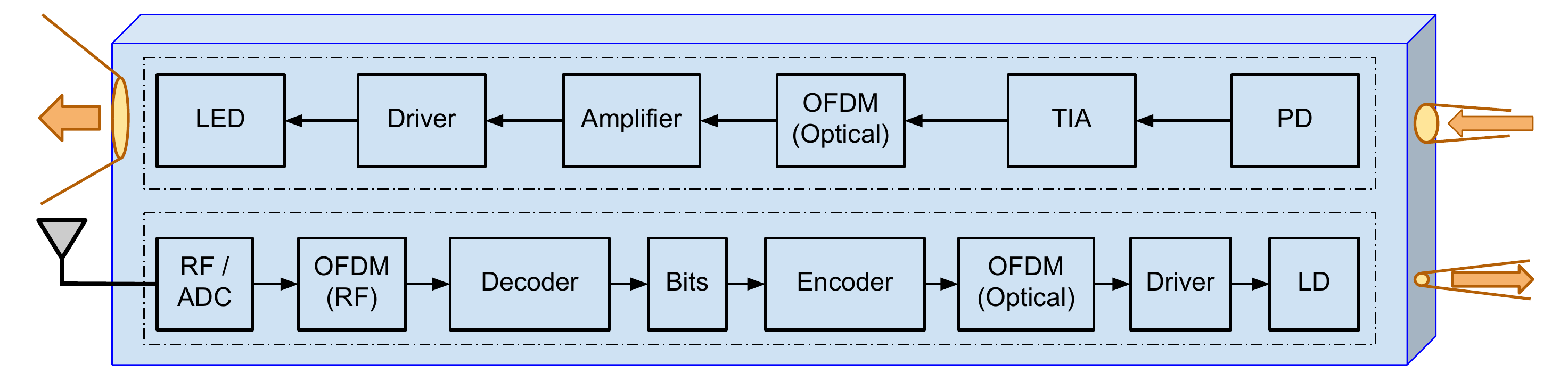}
		\caption{The VLC AP block diagram.}
		\label{Fig:4011}
	\end{subfigure}
	\caption{Proposed VLC-based W-OWPAN, with the structure of the VLC AP, its block diagram, and the amplification circuit.}
	\label{Fig:4}
\end{figure}
\par In Fig.~\ref{Fig:40}, we propose an architecture for VLC-based W-OWPAN. It highlights the duplex link required between UDs and APs (indoor), and between APs (outdoor). The system links two VLC-based OWPANs (OWPAN \textit{A} and OWPAN \textit{B}) in which nodes (A$_1$ and A$_2$; B$_1$ and B$_2$) are respectively connected to APs (VLC AP$_1$ and VLC AP$_2$) in an aggregate VLC-based network. The VLC AP$_1$-to-VLC AP$_2$ link is a laser beam connection, whereas the VLC AP-to-UD link has LED-based downlink traffic and is also used for illumination, and the uplink UD-to-AP is based on RF.   
\begin{figure}
	\centering
	\begin{subfigure}{0.48\textwidth}
		\includegraphics[width=\textwidth]{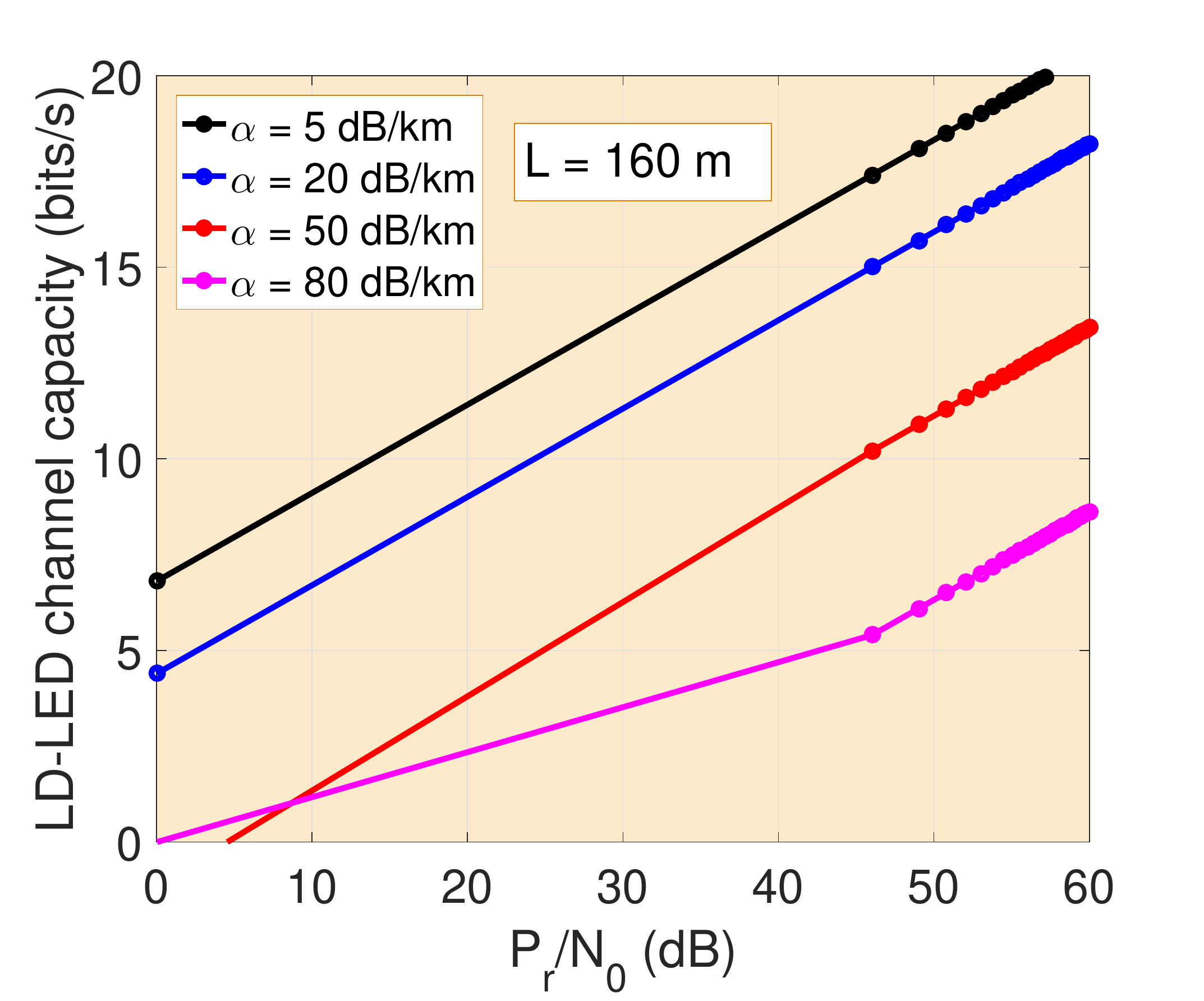}
		\caption{Capacity of a simplex downlink of a W-OWPAN versus $P_r/N_0$, for different values of the attenuation factor, $\alpha$.}
		\label{Fig:41}
	\end{subfigure}
	\begin{subfigure}{0.48\textwidth}
		\includegraphics[width=\textwidth]{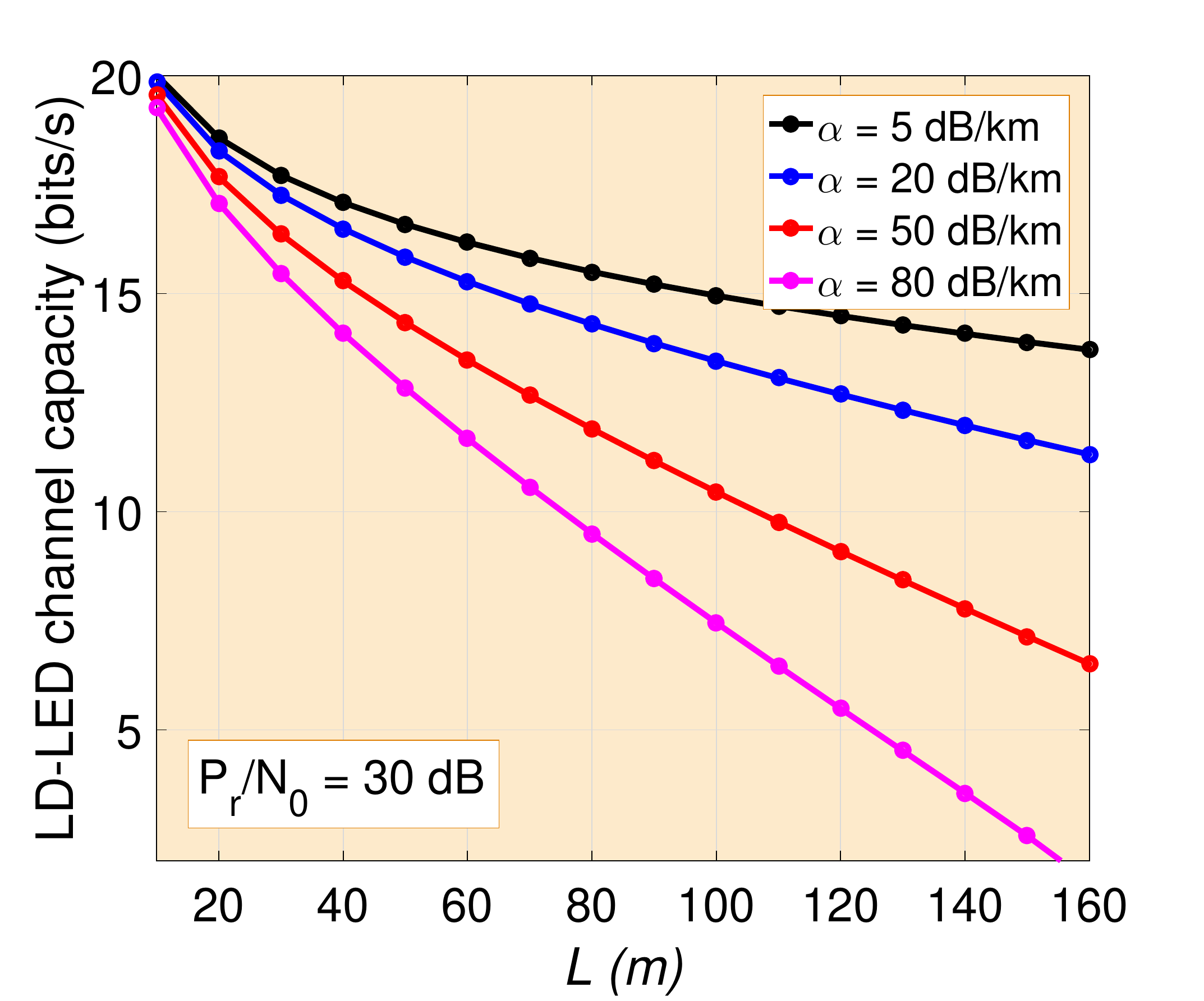}
		\caption{Capacity of a simplex downlink of a W-OWPAN versus $L$, for different values of the attenuation factor, $\alpha$.}
		\label{Fig:42}
	\end{subfigure}
	\caption{Channel capacity of VLC-based W-OWPAN.}
	\label{Fig:5}
\end{figure}
\subsection{Network and VLC AP Structures}

\begin{table}
	\begin{center}
		\caption{Parameters used in the analysis.}
		\label{tab:table0}
		\small
		%		\resizebox{\textwidth}{!}{%
		\begin{tabular}{p{7cm} p{2cm} p{3.5cm} p{2cm}} % <-- Alignments: 1st column left, 2nd middle and 3rd right, with vertical lines in between
			\rowcolor{cyan}
			\toprule
			Parameter & Symbol & Value & Unit \\
			\midrule
			%\rowcolor{Gray}
			Attenuation coefficient & $ \alpha $ & 5, 20, 50, 80 & dB/km  \\ 
			\midrule
				%\rowcolor{Gray}
			Distance between OWPANs & \textit{L} & 160 & m  \\ 
			\midrule
			Laser detector effective area & $A_c$ & 100 & mm$^2$ \\ 
			%	\rowcolor{Gray}
			\midrule
			Small angle divergence & $\theta$ & 8.38$\times$10$^{-7}$ & rads \\ 
			\midrule
			Room area & $A_{r}$ & 25 & m$^2$ \\ 
				%\rowcolor{Gray}{10%}
			\midrule
			Photodetector area & $A_{a}$ & 26 & mm$^2$ \\ 
			\midrule
			Average walls' reflectivity & $\rho$ & 0.7 & N/A \\ 
			%	\rowcolor{Gray}
			\midrule
			Angle of incidence & $\phi$ & 1.2217 & rads\\ 
			\midrule
			Half-intensity beam angle & $\phi_{1/2}$ & 0.5236 & rads \\ 
			%	\rowcolor{Gray}
			\midrule
			Distance between LED and PD & \textit{d} & 2.5 & m \\ 
			\midrule
			Angle of iradiance & $\Phi$ & 1.7453 & rads \\ 
			%	\rowcolor{Gray}
			\midrule
			Frequency & $f$ & $10$ & MHz \\ 
			\midrule
			Cutoff frequency at 3 dB & $f_0$ & 1.7111 & MHz \\ 
			%	\rowcolor{Gray}
			\midrule
			Delay of the line-of-sight link & $t_{1}$ & 0.01 & ns \\ 
			\midrule
			Delay of non line-of-sight link & $t_{2}$ & 0.03 & ns \\ 
			%	\rowcolor{Gray}
			\midrule
			Responsivity of the laser detector & $\lambda_{1}$ & 0.8 & N/A \\ 
			\midrule
			Responsivity of the PD & $\lambda_{1}$ & 0.8 & N/A \\
			%	\rowcolor{Gray} 
			\midrule
			Received power/noise power & $P_{r}/N_0$ & 30 & dB \\ 
			\midrule
			Amplification coefficient & $g_{0}$ & 1 & N/A \\ 
			%	\rowcolor{Gray} 
			\midrule
			Beam natural waist & $\omega_0$ & 0.588 & mm \\ 
			\midrule
			Wavelength & $\lambda$ & 1550 & nm \\ 
			\bottomrule 
		\end{tabular}
	\end{center}
\end{table}
The VLC AP, whose structure is shown in Fig.~\ref{Fig:400}, performs four main operations: one VLC detection, one RF detection and two VLC transmissions. Thus, each AP consists of a VLC receiver, an RF receiver, and two VLC transmitters. The VLC transmitters include two light sources. The first light source, which is used for both lighting and communication [the downlink (a) in Fig.~\ref{Fig:40}] is made up of LEDs, and utilizes the P2MP topology. It represents the message source for the optical connection between VLC AP and UDs. The second light source is an LD, which produces a light with a narrow beam due to its small divergence angle [link (c) in Fig.~\ref{Fig:40}]. This light source connects the two VLC APs (P2P) and is not used for illumination. The PD included in each VLC AP is used to detect the light produced by the opposite counterpart. The UDs are equipped with a PD, which is in charge of detecting the optical signal from link (a) and converting it into an electric current, and an RF transmitter which connects UDs to the VLC AP through link (b). Here, the RF receiver included in the VLC AP detects the uplink signal. \par For a complete connection between the two sub-networks, OWPAN \textit{A} and OWPAN \textit{B}, the overall connection is RF (b)-LD (c)-LED (a), in both directions. The first relaying scenario, which is RF (b)-LD (c) (uplink traffic), is based on decode-and-forward strategy, because the signals over both channels are symmetric (positive and negative) for RF and strictly positive for VLC. In this case, a direct signal conversion using the amplify-and-forward strategy, would result in an important loss of data. The second relaying scenario, which is LD (c)-LED (a) (downlink traffic), is built on the amplify-and-forward strategy, as it provides the system with cost-effective advantages. Here, the relay uses the circuit diagram shown in Fig.~\ref{Fig:401}. The related amplification gain (see Figs.~\ref{Fig:400} and \ref{Fig:401}), helps to mitigate the attenuation experienced by the optical signal over link (c). \par The VLC AP, whose block diagram is shown in Fig.~\ref{Fig:4011}, is made up of two parallel equipment to achieve full signal re-transmission. Data to be sent from any UD of OWPAN \textit{A} is firstly treated using the orthogonal frequency division multiplexing (OFDM) principle, for example. Afterwards, the obtained signal is transmitted using the RF channel (uplink). The resulting signal is collected from the RF channel by the RF antenna (in VLC AP$_1$) and digitized. The obtained digital signal is treated and the resulting bits are decoded. These bits are re-encoded, and the packets are converted into optical OFDM signals. Through an LD-driver, the latter is forwarded to an LD for re-transmission. At VLC AP$_2$, a PD equipped with a lens to concentrate the beam on the receiving area and a transconductance amplifier converts the laser light into an electrical signal. The latter is then amplified and forwarded to a conventional high-power LED via an analog LED-driver, for simultaneous lighting and communication whithin OWPAN \textit{B}. The LED broadcasts a light containing optical OFDM signals, which are collected by the PDs of the UDs. Finally, this received signal is demodulated, decoded and the resulting bits are sent to the recipient. 
\subsection{W-OWPAN Channel Capacity}
The link evaluation of the proposed W-OWPAN is fundamentally related to the overall channel behavior, which is a combination of the channels of links (a), (b) and (c), as the overall link is a cascaded combination of the three links. Thus, the performance of the system is dictated by the worst channel. Since link (b), which is RF-based, has been extensively studied in the literature, the channels analyzed are only those of links (a) and (c). Link (a) is the indoor optical part of the system, and its channel is a stochastic beam based on the diffuse topology. Its gain mainly depends on the geometrical parameters of the room and those of the beam. Link (c), the outdoor section of the system, is a narrow beam due to the small divergence angle and is based on the P2P topology. Its channel depends on: (\textit{i}) the attenuation of the laser power through the atmosphere, which is deterministic and described by the exponential Beers-Lambert law; (\textit{ii}) the atmospheric turbulence; and (\textit{iii}) the pointing error also modeled as Beckmann misalignment. Note that atmospheric turbulence and pointing errors are random and their distributions vary. Apart from these, the signal-to-noise ratio takes into account the laser detector effective area, the attenuation coefficient, $\alpha$, the input power, the distance between OWPANs, $L$, the beam angle, the transmission bandwidth, and the noise power. The maximum rate related to link (a) can be calculated based on its own signal-to-noise ratio, which fundamentally depends on the PD's effective area, the receiver field-of-view, the light incident angle, the Lambertian order, the transmit power, the transmission bandwidth, the amplification gain, the distance between VLC APs and UDs, and the noises over the channel. The link between OWPAN-\textit{A} and OWPAN-\textit{B} is established without significant turbulence effects and pointing error. \par The analytical results shown in Fig.~\ref{Fig:5} are obtained using the parameters displayed in Table~\ref{tab:table0}. These values are commonly used in the literature in practical implementation of both free space optical and indoor VLC. Taking into account the ratio of received signal power to the received noise power, \textit{$P_r/N_0$}, it is clear that the channel capacity increases as $P_r/N_0$ increases regardless of the value of $\alpha$ (Fig.~\ref{Fig:41}). This capacity decreases as the outdoor distance, $ L $, increases (Fig.~\ref{Fig:42}). For low values of $\alpha$, this decrement is slow, while it exhibits a fast drop for higher values of $\alpha$. Although some outdoor VLC impairment factors, such as turbulence, scintillation and pointing error, are not taken into account here, these results give a good indication on the harshness of the overall channel, which is mostly affected by the outdoor part of the network.
\section{Standardization Efforts}
\begin{table}[!htbp]
	\centering
	\caption{IEEE 802.15.7/D3a: Summary of the operating modes for PHY I to VI \cite{8569020}.}
	\label{tab:table1}%*7c
	\small
	\resizebox{\textwidth}{!}{%
		\begin{tabular}{p{2.5cm} p{3.2cm} p{4cm} p{2.7cm} p{2.7cm} p{4.2cm} p{3.3cm}}
			\rowcolor{cyan}
			\toprule
			\textbf{Mod. scheme} & \textbf{Line Code} & \textbf{Clock Rate} & \multicolumn{2}{c}{\textbf{Forward Error Correction}} & \textbf{Data Rate} & \textbf{Std. Version}\\
			\midrule
			%	\rowcolor{LightCyan}
			{}   & {}   & {}  & \textbf{Outer Code}   & \textbf{Inner Code} & {} & {}\\
			\midrule
			%	\rowcolor{Gray}
			\multicolumn{7}{c}{|\textbf{PHY I}|} \\
			\midrule
			OOK   &  Manchester & 200 kHz   & RS  & CC & 11.67 kbps to 100 kbps & $ \boxdot $, $ \boxtimes $/D2a/D3/D3a\\
			VPPM   &  4B6B & 400 kHz   & RS  & -.- & 35.56 kbps to 266.6 kbps & $ \boxdot $, $ \boxtimes $/D2a/D3/D3a\\
			\midrule
			%	\rowcolor{Gray}
			\multicolumn{7}{c}{|\textbf{PHY II}|} \\
			\midrule
			VPPM   &  4B6B & 3.75 MHz/7.5 MHz  & \multicolumn{2}{c}{RS} & 1.25 Mbps to 5 Mbps &$ \boxdot $, $ \boxtimes $/D2a/D3/D3a \\
			OOK   &  8B10B & 15 MHz to 120 MHz  & \multicolumn{2}{c}{RS} & 6 Mbps to 96 Mbps &$ \boxdot $, $ \boxtimes $/D2a/D3/D3a \\
			\midrule
			%	\rowcolor{Gray}
			\multicolumn{7}{c}{|\textbf{PHY III}|} \\
			\midrule
			CSK   &  -.- & 12 MHz/24 MHz  & \multicolumn{2}{c}{RS} & 1.25 Mbps to 5 Mbps & $ \boxdot $, $ \boxtimes $/D2a/D3/D3a\\
			OOK   &  8B10B & 15 MHz to 120 MHz  & \multicolumn{2}{c}{RS} & 12 Mbps to 96 Mbps & $ \boxtimes $/D3/D3a \\
			\midrule
			%	\rowcolor{Gray}
			\multicolumn{7}{c}{|\textbf{PHY IV}|} \\
			\midrule
			UFSOOK   &  -.- & Multiframe rate  & \multicolumn{2}{c}{MIMO path dependent} & 10 bps & $ \boxtimes $/D3/D3a\\
			Twinkle VPPM   &  -.- & 4x bit rate  & \multicolumn{2}{c}{RS} & 4 kbps & $ \boxtimes $/D3/D3a\\
			S2-PSK   &  Half-rate code & 10 Hz  & \multicolumn{2}{c}{Temporal error correction} & 5 kbps & $ \boxtimes $/D3/D3a\\
			HS-PSK   &  Half-rate code & 10 kHz  & \multicolumn{2}{c}{RS} & 22 kbps & $ \boxtimes $/D3/D3a\\
			Offset VPPM   &  -.- & 25 Hz  & \multicolumn{2}{c}{RS} & 18 bps & $ \boxtimes $/D3/D3a\\
			\midrule
			%	\rowcolor{Gray}
			\multicolumn{7}{c}{|\textbf{PHY V}|} \\
			\midrule
			RS-FSK   &  -.- & 30 Hz  & \multicolumn{2}{c}{XOR FEC} & 120 bps & $ \boxtimes $/D3/D3a\\
			C-OOK   & Manchester/4B6B & 2.2 kHz/4.4 kHz  & Hamming code & Optional/RS & 400 bps & $ \boxtimes $/D3/D3a\\
			CM-FSK   & -.- & 10 Hz  & -.- & Optional & 60 bps & $ \boxtimes $/D3/D3a\\
			MPM   &  -.- & 12.5 kHz  & \multicolumn{2}{c}{Temporal error correction} & 7.51 bps & $ \boxtimes $/D3/D3a\\
			\midrule
			%	\rowcolor{Gray}
			\multicolumn{7}{c}{|\textbf{PHY VI}|} \\
			\midrule
			A-QL   &  -.- & 10 Hz  & RS & CC & 5.54 kbps & $ \boxtimes $/D3/D3a\\
			HA-QL   & Half-rate code & 10 Hz & RS & CC & 140 bps & $ \boxtimes $/D3/D3a\\
			VTASC   & -.- & 30 Hz  &  \multicolumn{2}{c}{RS} & 512 kbps & $ \boxtimes $/D3/D3a\\
			SS2DC   &  -.- & 30 Hz  &  \multicolumn{2}{c}{RS} & 368 kbps & $ \boxtimes $/D3/D3a\\
			IDE-MPSK BLEND   & -.- & 30 Hz  &  \multicolumn{2}{c}{RS} & 32 kbps & $ \boxtimes $/D3/D3a\\
			IDE-WM   &  -.- & 30 Hz  &  \multicolumn{2}{c}{RS} & 256 kbps & $ \boxtimes $/D3/D3a\\
			\bottomrule \\
			\multicolumn{7}{c}{$\boxdot $ = IEEE Std 802.15.7-2011; $\boxtimes $ = IEEE P802.15.7; OOK = on-off keying; VPPM = variable pulse position modulation; CC = convolutional encoder;} \\
			\multicolumn{7}{c}{UFSOOK = undersampled frequency shift OOK; PSK = phase-shift-keying; S2-PSK = spatial 2-PSK; C-OOK = camera OOK;} \\
			\multicolumn{7}{c}{CM-FSK = camera M-ary frequency-shift-keying; VTASC = variable transparent amplitude-shape-color; SS2DC = sequential scalable 2D code;}\\ 
			\multicolumn{7}{c}{A-QL = asynchronous quick link; HS-PSK = hybrid spatial phase-shift-keying; HA-QL = hidden asynchronous quick link.} \\
			\multicolumn{7}{c}{-.- no line code or no forward error correction code is proposed.} \\
	\end{tabular}}
\end{table}
The deployment of VLC technology and efficient implementation of OWPANs require rules and regulations, which are still in the development phase. A few drafts have recently been prepared, each having specific focuses. Examples are those proposed by the Visible Light Communication Consortium and IEEE. Among them, IEEE provides the major contribution because its proposition is broader and looks at most aspects of the VLC technology. In the sequel, we focus on the efforts of IEEE, which is led by the IEEE 802.15.7 Task Group. It proposes a layer-based architecture which follows the IEEE 802.15.4 layout based on TCP/IP. Their efforts focus on media access control (MAC) and PHY layers because the upper layers are similar to the other networking standards, such as IEEE 802.11. An overview of the IEEE 802.15.7 MAC and PHY layers is discussed next. 
\subsection{MAC Layer Frame}
The IEEE 802.15.7 standard proposes a MAC frame which consists of three main parts, namely the MAC header, payload, and footer. It depends on the targeted PHY layer (PHY I to VI), and the network topology (P2P, P2MP, OWPAN, etc.). The standard supports three main MAC network types, which are P2P, star, and broadcast \cite{8569020}. P2P uses the configuration in Scenario (1) of Fig.~\ref{Fig:10}, while star and broadcast use Scenarios (2) and (3), and are based on the P2MP topology. In each of these network categories, the nodes have a unique 64-bit address that is used for communication within the network, which in general is an OWPAN. The P2P topology may be a duplex transmission that uses the configuration illustrated in Fig.~\ref{Fig:21} based on an extra-narrow beam, while the other types use the diffuse link shown in Fig.~\ref{Fig:22}. \par To achieve transmission in OWPAN using the P2MP topology, the IEEE 802.15.7 MAC layer utilizes a super-frame format. It is used to group and align channels in order to determine channel locations in the received stream, and is defined by the coordinator (VLC AP). The latter is the owner of the super-frame, and consequently, allocates guaranteed time slots and forms the contention free period.
\subsection{Physical Layer}
The IEEE 802.15.7 PHY layer has evolved within the original and actual version of the standard. Compared to the original version of IEEE 802.15.7, which has three categories of physical layers (PHY I, PHY II, and PHY III), the newer version, IEEE 802.15.7/D3a, divides the VLC PHY layer into six categories, i.e., PHY I to VI. Each of them is characterized by particular specifications related to the modulation scheme, forward error correction techniques, data rate, to mention only a few parameters. PHY I is dedicated to data rates between 11.67 kbps and 266.6 kbps. It is divided into two sections: 11.67 kbps to 100 kbps and 35.56 kbps to 266.6 kbps depending on the optical clock rate used. Hence, the choice of optical clock rate determines if the header is transmitted with 11.67 kbps or 35.56 kbps. 11.67 kbps is used for an optical clock rate of 200 kHz. A Manchester code is employed as line code, and Reed Solomon and convolutional codes are used for error correction. They are combined with the on-off keying, selected as the modulation scheme. 35.56 kbps is selected to transmit the header for an optical clock rate of 400 kHz. Here, variable pulse position is the modulation of choice with 4B6B line code and Reed Solomon as an outer code. A full description of the six PHY layers proposed in IEEE 802.15.7/D3a is provided in Table~\ref{tab:table1}.
\section{Challenges related to VLC-based Networking}
In light of the above discussion, it is clear that one of the most important challenges of VLC-based networks practical deployment resides in the integration of the technology used in the uplink. It is also worth underlining the dual-use of the light source, as a telecommunication antenna and illumination device, which requires efficient algorithms to meet both lighting and communication requirements. \par In case of mobile UDs, the handover mechanism represents another challenge if several VLC APs are used. It is important to note that, since the propagation of the transmitted signal in VLC technology is different compared to RF, the handover techniques exploited for RF systems cannot be directly applied to VLC-based networks \cite{8101446}. \par In aggregate VLC-based networks involving technologies such as RF, fiber optic, power line communications, and free space optical, the message frame compatibility is a critical issue that must be addressed. VLC systems must be compatible with the existing schemes. \par Relayed networks based on the amplify-and-forward strategy are cost-effective compared to other schemes such as decode-and-forward, because the signal is just amplified and forwarded without any processing. For this reason, the amplify-and-forward scenario should be the primary choice. The modulation schemes exploited in the first link should be similar or compatible with that of the second link, for an efficient implementation of the bridge between the channels. In this case, finding the correct combination between modulation schemes to be used in different technologies represents another challenge. \par Note that these challenges, related to single channel systems, increase with the number of channels in multi-channel networks. \par More importantly, high capacity and throughput represent two challenges in VLC-based networking due to the bandwidth limitations of LEDs and LDs. However, solving this specific drawback is the responsibility of not only the networking designers, but also that of the LEDs/LDs manufacturers. Any solution to this design aspect should maintain compatibility between data transmission and illumination functions.
\section{Research Opportunities for VLC-based Networks}
The proof of the VLC concept has now been well established \cite{7402263,7224733}. A few companies such as Pure Li-Fi have started to manufacture VLC APs and optical components for mobile devices such as laptops and mobile phones. In a near future, VLC equipment will be extended to other VLC systems including standalone and aggregate, relayed and non-relayed, simplex and duplex, homogeneous and heterogeneous, and multi-channel VLC-based systems. Several applications are targeted including indoor positioning, internet broadcasting, massive data transfer between two VLC nodes, underwater communications, intelligent transportation systems, and health-care applications. \par There are considerable opportunities for the design and implementation of VLC-based networks. Research is needed to explore VLC-based networking in different VLC-based network types discussed in this article. These new efforts will be associated not only with research in telecommunication, but also with improvement happening in the lighting industry, which will lead to undeniable enhancement of both lighting and communication performances. \par The heterogeneous network will require more studies and analysis before integration, as the technologies and protocols in place need to be compatible. Synchronization between frequencies in some cases needs to be handled carefully. \par More experimental results and practical deployments are required to corroborate results obtained by theoretical analysis. This applies to all enumerated VLC-based network types and represents a great deal of research opportunity for standalone, aggregate, homogeneous, heterogeneous, duplex, and relayed networks using VLC technology. \par Most research results in VLC technology are related to Li-Fi, intelligent transportation systems, indoor positioning and some cooperative systems. VLC can be associated with power line communications, Wi-Fi and free space optical to form cost-effective communication systems. Emerging VLC applications such as massive data transfer between two VLC nodes and those in the underwater environment need to be investigated. VLC can mimic technologies such as Bluetooth to transfer files between two nodes. This has not been investigated yet. Potential underwater applications of VLC that deserve to be explored include surveillance, climate change, and oceanic fauna monitoring. Channel modeling, network protocols, design and practical implementation issues of these applications represent potential research directions.
\section{Conclusion}
This article has elaborated on VLC-based networking through possible connectivity configurations, along with applicable topologies and technologies for the return path in duplex transmissions. The protocol models are highlighted and reveal that VLC-based networks use the layered protocol with differences at both MAC and PHY layers compared to conventional networks. Specifications for these two layers are provided for the IEEE 802.15.7/D3a standard in terms of modulation and forward error correction schemes. An architecture of a VLC-based network for sharing data between two OWPANs, linked using a duplex outdoor VLC backbone subsystem, is also proposed and discussed, and the related link capacity is presented. Finally, research opportunities toward the utilization of VLC technology in OWPAN and its integration into existing conventional networks is outlined.

\bibliographystyle{unsrt}  
\bibliography{mag1}

\end{document}